\documentclass[twoside,12pt]{article}
\usepackage{epsfig}

\def\lab{\label}
\def\lan{\langle}
\def\lf{\left}

\def\ran{\rangle}

\def\ri{\right}

\def\al{\alpha}

\def\te{\theta}

\def\si{\sigma}

\def\1{{_{1}}}\def\2{{_{2}}}
\newcommand{\bea}{\begin{eqnarray}}\newcommand{\eea}{\end{eqnarray}}
\newcommand{\beaa}{\begin{eqnarray}}\newcommand{\eeaa}{\end{eqnarray}}
\newcommand{\ba}{\begin{array}}\newcommand{\ea}{\end{array}}
\newcommand{\bit}{\begin{itemize}}\newcommand{\eit}{\end{itemize}}
\newcommand{\ben}{\begin{enumerate}}\newcommand{\een}{\end{enumerate}}
\begin{document}

\title{\vspace{1cm} Particle mixing, flavor condensate and dark energy}

\author{Massimo Blasone, Antonio Capolupo and Giuseppe Vitiello\\
\\
Dipartimento di Matematica e Informatica and INFN
\\
Universit´a  di  Salerno,  Fisciano  (SA)-84084,  Italy.}

\maketitle

\begin{abstract}

The mixing of neutrinos and quarks generate a  vacuum condensate that, at the present epoch,
behaves as a cosmological constant. The value of
the dark energy is constrained today by the very small breaking of the Lorentz invariance.

\end{abstract}

The accelerated expansion of the universe today observed \cite{SCP}-\cite{WMAP-Five}
is explained with the hypothesis that almost $70\%$ of the energy content of the universe
is due to an homogeneous fluid that has negative pressure, called dark energy. The nature
of this energy component remains unknown.

Here we report on  recent results \cite{Capolupo:2008rz} according to which
the vacuum condensate generated by the particle mixing
\cite{Capolupo:2006et}-\cite{Blasone:2004yh}
could explain the dark energy of the universe.
In particular, at the present epoch, the small breaking of the Lorentz invariance of the flavor vacuum
forces the small value of the dark energy \cite{Capolupo:2008rz}.

We briefly present the quantum field theory formalism for mixed fields
\cite{BV95}-\cite{Blasone:2005ae} (for a detailed review see \cite{Capolupo:2004av}).
We consider the mixing among three generations of Dirac fields and
show the contribution to the dark energy given,
at the present epoch, by particle  mixing  \cite{Capolupo:2008rz}.

The mixing transformations are:
$ \Psi_f(x) \, = {\cal U} \, \Psi_m
(x)$,
where ${\cal U}$ is the CKM matrix
and $\Psi_m^T=(\psi_1,\psi_2,\psi_3)$ are
the fields with definite masses $m_{1} \neq m_{2} \neq m_{3}$.
By means of  the mixing generator $G_{\bf \te}(t)$ \cite{Capolupo:2008rz,yBCV02,Capolupo:2004av},
the mixing relations can be expressed as
$\psi_{\si}^{\al}(x)\equiv G^{-1}_{\bf \te}(t) \,
\psi_{i}^{\al}(x)\, G_{\bf \te}(t), $ where $(\si,i)=(A,1), (B,2),
(C,3)$ with $A$, $B$, $C$  lepton $(e, \mu, \tau)$
 or flavor $(d, s, b)$ indices.
The flavor vacuum is given by $ |0(t) \rangle_{f} = G^{-1}_{\bf \te}(t)\;
|0 \rangle_{m}\,, $
where $|0\rangle_{m}$ is the vacuum for fields with definite
masses annihilated by $\alpha ^{r}_{{\bf k},i}$ and $ \beta ^{r
}_{{\bf k},i}$, $ i=1,2,3 \;, \;r=1,2$.
The vacuum $|0\rangle_{f}$ is annihilated by the operators:
$
\alpha _{{\bf k},\sigma}^{r}(t) \equiv G^{-1}_{\bf \te}(t)\;\alpha
_{{\bf k},i}^{r}(t)\;G_{\bf \te}(t)\,,$ and
$\beta_{{\bf k},\sigma}^{r}(t) \equiv G^{-1}_{\bf \te}(t)\;\beta
_{{\bf k},i}^{r}(t)\;G_{\bf \te}(t)\,$.
In the infinite volume limit  $|0(t) \rangle_{f}$ is unitarily
inequivalent to  $|0 \rangle_{m}$ \cite{BV95}, \cite{hannabuss}.
Moreover, $|0(t) \rangle_{f}$ is a coherent condensate of particles
 whose numbers, in the reference frame such that
${\bf k}=(0,0,|{\bf k}|)$,  are \cite{yBCV02}:
\bea \lab{V1}
{\cal N}^{\bf k}_1\, = \,_{f}\langle0(t)| \alpha_{{\bf k},1}^{r \dag} \alpha_{{\bf k},1}^{r} |0(t)\ran_{f} \,
= \,_{f}\langle0(t)|\beta_{{\bf k},1}^{r \dag} \beta_{{\bf k},1}^{r}|0(t)\ran_{f}\,
 = \, s^{2}_{12}c^{2}_{13}\,|V^{{\bf k}}_{12}|^{2}+ s^{2}_{13}\,|V^{{\bf k}}_{13}|^{2}\,,
\eea
and similar relations for ${\cal N}^{\bf k}_2$, ${\cal N}^{\bf k}_3$.
In Eq.(\ref{V1}),  $V^{{\bf k}}_{ij}$ are the Bogoliubov coefficients entering the mixing transformations
(see  Refs.\cite{Capolupo:2008rz,yBCV02,Capolupo:2004av}).

The condensate
 due to particle mixing behaves as a perfect fluid \cite{Capolupo:2008rz}.
Indeed, its energy momentum tensor density:
\bea \label{T-Cond}
{\cal T}_{\mu\nu}^{cond}(x)={}_{f}\lan 0(t) |:{\cal T}_{\mu\nu}(x):| 0(t)\ran_{f}\,,
\eea
can be written as
$
{\cal T}_{\mu\nu}^{cond}\,
 = \, diag
({\cal T}_{00}^{cond}\,,{\cal T}_{11}^{cond}\,,{\cal T}_{22}^{cond}\,
,{\cal T}_{33}^{cond}\,)\,.
$
In Eq.(\ref{T-Cond}) $:{\cal T}_{\mu\nu}(x):$ denotes the energy-momentum tensor density
 for the fermion fields $\psi_i$, $i=1,2,3$ in the Minkowski metric.

The tensor ${\cal T}_{\mu\nu}(x) $ can be written as
\bea
 :{\cal T}_{\mu\nu}(x): & = &
 : \Sigma_{\mu\nu}(x):\,+\,:{\cal V}_{\mu\nu}(x):
 \eea
where
\bea
:\Sigma_{\mu\nu}(x): & = &
:\lf\{\frac{i}{2}\left({\bar \Psi}_{m}(x)\gamma_{\mu}
\stackrel{\leftrightarrow}{\partial}_{\nu} \Psi_{m}(x)\right)
- \eta_{\mu\nu} \lf[\frac{i}{2} {\bar \Psi}_{m}(x)  \gamma^{\alpha}
\stackrel{\leftrightarrow}{\partial}_{\alpha}  \Psi_{m}(x) \ri]\ri\}: \;,
\\
:{\cal V}_{\mu\nu}(x): & = &
 \eta_{\mu\nu} : \lf[ {\bar \Psi}_{m}(x) \, \textsf{M}_d \,  \Psi_{m}(x) \ri]: \;,
 \eea
 $\textsf{M}_d= diag(m_{1}, m_{2}, m_{3})$, $\Psi_{m} = (\psi_1, \psi_2, \psi_3)^{T}$
 and $\eta_{\mu\nu} = diag (1,-1,-1,-1)$.

The contributions given by
particle mixing to the vacuum energy density $\rho_{mix}$
and to the vacuum pressure $ p_{mix}$ are respectively:
 \bea\label{ro}
\rho_{mix} & \equiv & \frac{1}{ V}\; \eta^{00}\; \int d^{3}x \;{\cal T}_{00}^{cond}(x) \, =\,
\frac{2}{\pi} \sum_{i}\,  \int d k \, k^{2}\,
\omega_{k,i}\;  {\cal N}^{\bf k}_i\,,
 \\
 \label{p-mix}
p_{mix} & \equiv &  -\frac{1}{ V}\; \eta^{jj} \; \int d^{3}x \;{\cal T}_{jj}^{cond}(x) \, =\,
 \frac{2}{3 \pi} \sum_{i}\,\int d k \, k^{2} \, \frac{k^{2}}{\;\omega_{k,i}}\;
 {\cal N}^{\bf k}_i\,,
 \eea
(no summation on $j$ is intended).

In particular, in the present epoch, the very small breaking of the Lorentz invariance
\cite{WMAP-Five}, imposes that
 ${\cal T}^{cond}_{\mu\nu}(x)$
is space-time independent.
Then, the kinematical part $\Sigma^{cond}_{\mu\nu}$
of ${\cal T}^{cond}_{\mu\nu}$ is negligible \cite{Capolupo:2008rz}:
$
\Sigma^{cond}_{\mu\nu}\,=\;{}_{f}\lan
0(t) |:\Sigma_{\mu\nu}(x):| 0(t)\ran_{f}\,\simeq 0
$
and ${\cal T}_{\mu\nu}^{cond}$ is given today by:
\bea\label{Tcond} {\cal T}_{\mu\nu}^{cond} \simeq \;{}_{f}\lan
0(t) |:{\cal V}_{\mu\nu}(x):| 0(t)\ran_{f}\, =\,
\eta_{\mu\nu}\;{}_{f}\lan
0(t) |:{\bar \Psi}_{m}(x)\;\textsf{M}_d\;\Psi_{m}(x):| 0(t)\ran_{f}\,.
 \eea
Thus, we have:
\bea\label{Tcond1}
 diag
(\rho_{mix}\,,p_{mix}\,,p_{mix}\,,p_{mix})
\,=\, \eta_{\mu\nu}\;\sum_{i}m_{i}\int \frac{d^{3}x}{(2\pi)^3}\;{}_{f}\lan
0 |:\bar{\psi }_{i}(x)\psi_{i}(x):| 0\ran_{f}\,.
 \eea
Eq.(\ref{Tcond1}) implies that, at the present epoch, the vacuum condensate generated from particle
mixing has the state equation characteristic of the cosmological constant: $\rho_{mix} \simeq -p_{mix}$ \cite{Capolupo:2006et}. The adiabatic index is then
$
w_{mix}\; = \; p_{mix}/ \rho_{mix}\; \simeq \; - 1\;,
$
where the contribution $\rho_{mix}$ is \cite{Capolupo:2008rz}:
 \bea \label{cost1}
\rho_{mix} & \simeq & \frac{2}{\pi} \sum_{i} \int_{0}^{K} dk \, k^{2}\;
\frac{m_{i}^{2}}{\omega_{k,i}} \;{\cal N}^{\bf k}_i\,.
 \eea
$K$ is the cut-off on the momenta.

The integral (\ref{cost1}) diverges in $K$ as $m_{i}^{4}\,\log\lf( 2K /m_{j}\ri)$,
 with $i,j = 1,2,3$ \cite{Capolupo:2006et}.
However, as shown in Ref.\cite{Capolupo:2008rz}, the value close to $-1$ of $w_{mix}$
 at the present epoch constrains the value of $K$ and consequently the
 value of $\rho_{mix}$.
We find the following results  by using different values of $w_{mix}$ close to  $-1$
\cite{Capolupo:2008rz}:

\vspace{3mm}
\textit{Neutrino mixing condensate contribution:}
%
$\rho^{\nu}_{mix} \sim  10^{-47} GeV^{4}$ for $-0.98\leq w^{\nu}_{mix} \leq -0.97\,.$
Such values are compatible with the estimated
upper bound of the dark energy and $w^{\nu}_{mix}$ is in agreement with
the constraint on the dark energy state equation \cite{WMAP-Five}.

Negligible contributions
of $\rho^{\nu}_{mix}$ are found for $w^{\nu}_{mix} < -0.98$. The results obtained are
dependent on the neutrino mass values one uses.

\vspace{3mm}
\textit{Quark mixing condensate contribution:}
%
$\rho^{q}_{mix} = 1.5 \times 10^{-47} GeV^{4}$
 for $w^{q}_{mix} =-1$ \cite{Capolupo:2008rz}.
Very small deviations from the value $w^{q}_{mix} = -1$
give rise to contributions of $\rho^{q}_{mix}$
that are beyond the accepted upper bound of the dark energy.

In conclusion, the vacuum condensate from particle mixing
provides a contribution to the dark energy which is compatible with the
estimated value of the cosmological constant. Such  value  is imposed by the small
 breaking of the Lorentz of the flavor vacuum at the present epoch.

\vspace{3mm}

It is worth to remark that it is possible to obtain the above results only
when one uses a field theoretical approach to the problem of particle mixing,
where a rich physical structure associated to the flavor vacuum emerges.
In this connection, it is important to note that the statements
appeared recently in some paper (Y.F.~Li and Q.Y.~Liu, { JHEP} { 0610}, 048 (2006))
are misleading and in fact wrong, as can be easily checked and will be shown in a forthcoming publication.
In particular, it can be proved exactly that in the present QFT formalism, no flavor charge violation arises,
in contrast to what it has been found to happen in the framework of the
conventional treatment \cite{Nishi:2008sc}.

\end{document}